\documentstyle[12pt]{article} 
\newcommand{\journal}[4]{{\em #1~}#2\,(19#3)\,#4;}

\newcommand{\jmp}{\journal {J. Math. Phys.}}

\newcommand{\cmp}{\journal {Comm. Math. Phys.}}

\newcommand{\np}{\journal {Nucl. Phys.}}  
\newcommand{\pl}{\journal {Phys. Lett.}}

\newcommand{\prep}{\journal {Phys. Rep.}}

\newcommand{\PRD}{\journal{\em Phys. Rev. D}} 

\makeatletter
\@addtoreset{equation}{section}
\makeatother

\renewcommand{\a}{\alpha}

\renewcommand{\d}{\delta}
\newcommand{\e}{\varepsilon}

\newcommand{\k}{\kappa}

\newcommand{\m}{\mu}
\newcommand{\n}{\nu}
\newcommand{\mn}{{\mu\nu}}

\renewcommand{\o}{\omega} 

\newcommand{\r}{\rho}
\newcommand{\s}{\sigma}

\newcommand{\BB}{{\cal B}}

\newcommand{\EE}{{\cal E}}
\newcommand{\FF}{{\cal F}}

\newcommand{\LL}{{\cal L}}
\newcommand{\MM}{{\cal M}}

\newcommand{\PP}{{\cal P}}

\newcommand{\RR}{{\cal R}}

\newcommand{\ZZ}{{\cal Z}}

\newcommand{\complex}{{\kern .1em {\raise .47ex
\hbox {$\scriptscriptstyle |$}}
    \kern -.4em {\rm C}}}
\newcommand{\real}{{{\rm I} \kern -.19em {\rm R}}}
\newcommand{\rational}{{\kern .1em {\raise .47ex
\hbox{$\scripscriptstyle |$}}
    \kern -.35em {\rm Q}}}
\renewcommand{\natural}{{\vrule height 1.6ex width
.05em depth 0ex \kern -.35em {\rm N}}}
\newcommand{\integers}{{\bf Z}}
\newcommand{\ipr}{\!\cdot\!}

\newcommand{\tr}{{\rm {Tr} \,}}

\renewcommand{\exp}{{\rm \ {exp}\,}}
\newcommand{\cb}{{\bar c}}

\newcommand{\half}{\frac 1 2}

\newcommand{\pa}{\partial}

\newcommand{\fud}[2]  {{\displaystyle{\frac{\delta #1}{\delta #2}}}}

\newcommand{\sla}{\raise.15ex\hbox{$/$}\kern -.57em}

\newcommand{\twiddle}{\lower.9ex\rlap{$\kern -.1em\scriptstyle\sim$}}

\newcommand{\equ}[1]{~(\ref{#1})}

\newcommand{\eq}{\begin{equation}}
\newcommand{\eqn}[1]{\label{#1}\end{equation}}
\newcommand{\eea}{\end{eqnarray}}
\newcommand{\eqa}{\begin{eqnarray}}
\newcommand{\eqan}[1]{\label{#1}\end{eqnarray}}
\newcommand{\ba}[1]{\begin{equation}\begin{array}{#1}}
\newcommand{\ea}[1]{\end{array}\label{#1}\end{equation}}
\newcommand{\eqac}{\begin{equation}\begin{array}{rcl}}
\newcommand{\eqacn}[1]{\end{array}\label{#1}\end{equation}}

\newcommand{\one}{{\bf 1}}

\newcommand{\norm}[1]{{|\!| {#1}|\! |}}
\newcommand{\ol}{\overline}

\newcommand{\ov}{\over}
\newcommand{\lf}{\left}
\newcommand{\rt}{\right}

\begin{document}
\def\ftoday{{\sl  \number\day \space\ifcase\month 
\or Janvier\or F\'evrier\or Mars\or avril\or Mai
\or Juin\or Juillet\or Ao\^ut\or Septembre\or Octobre
\or Novembre \or D\'ecembre\fi
\space  \number\year}}    
\titlepage
\hfill{\today} 
%
{
\begin{center}
{ \huge Topological Aspects of Gauge Fixing \\ \vskip 1ex 
Yang-Mills Theory on $S_4$}
\vspace{2ex}

{\Large Laurent Baulieu\footnote{e-mail~:
Baulieu@lpthe.jussieu.fr}}\\{\it\large LPTHE Paris VI-VII~\footnote{ Boite 126, Tour 16, 
1$^{\it er}$ \'etage,
        4 place Jussieu,
        F-75252 Paris CEDEX 05, FRANCE}
\footnote{Laboratoire associ\'e No. 280 au CNRS  et aux
 Universit\'e Pierre et Marie
Curie - Denis Diderot }\\and
\\{\it\large Yukawa  Institute for Theoretical Physics, Kyoto University~\footnote{ Kyoto University, Kyoto 606,
Japan}} }
\\{\Large \vskip.5ex and \\ \vskip.5ex
Alexander Rozenberg\footnote{e-mail~: sasha@acf2.nyu.edu} and Martin
Schaden\footnote{e-mail~: schaden@mafalda.physics.nyu.edu -- research
offers are welcome! \\ 
\indent{~~This research} was partly supported by the National Science
Foundation under grant no.~PHY93-18781}}\\ 
{\it\large Physics Department, New York University,\\ 4 Washington Place,
 New York, N.Y. 10003}
\end{center}
\vspace{4ex}

\begin{center}
\bf ABSTRACT
\end{center}
{For an $S_4$ space-time manifold  global aspects of gauge-fixing 
are investigated using the relation to
Topological Quantum Field Theory on the gauge group.  The
partition function of this TQFT is shown to compute the regularized Euler
character of a suitably defined  space of gauge
transformations. Topological properties of the
space of solutions to a covariant gauge conditon on the orbit of a
particular instanton are found using the $SO(5)$ isometry group of
the $S_4$ base manifold. We obtain that the Euler character of this
space differs from that of an orbit in the
topologically trivial sector.  This result implies that an orbit with 
Pontryagin number $\k=\pm1$ in covariant gauges on $S_4$ contributes
to physical correlation functions with a
different multiplicity factor due to the Gribov copies, than an orbit in
the trivial   $\k=0$ sector. Similar
topological arguments show that there is no contribution from the
topologically trivial sector to physical
correlation functions in gauges defined by a nondegenerate background
connection. We discuss possible physical implications of the global gauge
dependence of Yang-Mills theory.}  

PACS: 11.15.-q 11.15.Bt 11.15.Tk 11.15.Bx\hfill\break
NYU--TH--96/05/03\hfill HEP-TH/9607147
\vfill

\newpage
\def\be{\begin{eqnarray}}
\def\ee{\end{eqnarray}}
\def\nn{\nonumber}
\section{Introduction} 
Since the pioneering work of Gribov\cite{gr78} and Singer\cite{si78}
it is known that the gauge condition  
\eq
\FF[A]=0   
\eqn{cond}
generally does not  determine the representative connection
$A$ of a gauge orbit uniquely for non-abelian gauge groups such as
$SU(n)$. It is therefore of interest to study the space of gauge equivalent
solutions to\equ{cond}
\eq
\EE_\FF[A]=\{U: U(x)\in SU(n), \FF[A^U]=0\}
\eqn{ef}
on the gauge orbit $\{A^U\}$ with representative connection $A$
\eq
A^U=U^\dagger A U +U^\dagger d U.
\eqn{orb}

It was recently pointed out\cite{us96}, that gauge fixing of Yang-Mills
 theory defined on a compact spacetime manifold amounts to the
construction of a certain Topological Quantum Field Theory on
the space of gauge transformations, whose 
partition function computes some topological number of\equ{ef}. This
relation to TQFT enables us to investigate global aspects of the gauge
fixing procedure in greater detail than was previously possible.  We
will show in section~2 that the partition function of the TQFT
proposed in ref.\cite{us96} is proportional to  the Euler characteristic 
$\chi(\EE_{\pa\ipr A})$ of the moduli space 
\eq
\EE_{\pa\ipr A}[A]=\{U: U(x)\in SU(n), \pa\ipr A^U=0\}/SU(n)
\eqn{space}   
Note that the Euler character of
the space of solutions to the covariant (Landau) gauge condition
$\pa\ipr A^U=0$  vanishes due to the isometry with respect to right
multiplication by constant group elements. This isometry group has
been factored out in\equ{space} and it was shown\cite{us96} that
$\chi(\EE_{\pa\ipr A})=odd\neq 0$ in the vicinity of flat connections $A\simeq
U^\dagger d U $   for an $SU(2)$  gauge group defined on any compact
space-time manifold.  

Since the topological properties of the moduli space\equ{space} do not
change under continuous deformations, one is guaranteed that
$\chi(\EE_\FF[A])$ is constant within a topologically connected sector of
gauge orbits. The Euler character could however depend on: i) the
topological sector of the gauge orbit and/or ii) the gauge fixing
condition\equ{cond}.

The two possibilities are intimately related, since the
existence of topologically disconnected sectors in the orbit space
implies that background gauge conditions of the type
\eq
\FF[A]=D^B\ipr A-\pa\ipr B=0 
\eqn{background}
also cannot be deformed into each other for background connections
$B$ belonging to different topological sectors.

In this paper we will investigate both possibilities by considering
 an $SU(2)$ gauge theory defined on  $S_4$. Disconnected sectors of
the orbit space with different Pontryagin number 
$\k[A]\in\integers$ 
\eq
\frac{1}{4\pi^2}\int_{S_4} \tr F(A)\wedge F(A) = \k[A]\,;\qquad
F(A)= d A + A\wedge A    
\eqn{pontryagin} 
exist in this case. For a given gauge condition $\FF(A)$, the Euler
character  $\chi(\EE_\FF)$ of the moduli space
$\EE_\FF$ associated with an orbit can therefore in general depend on the
Pontryagin number of the orbit $\chi(\EE_\FF[A])=\chi_\FF(\k[A])$.  
Since an $S_4$ manifold has only trivial 1-cycles, the result presented in the
appendix of \cite{us96} actually implies that $\chi_\FF(\k[A]=0)=1$ for an
$SU(2)$ theory in any covariant gauge which can be deformed to Landau
gauge $\pa\ipr A=0$.  
In section~3 we discuss the moduli space $\EE_{cov}$ of such covariant
gauges for a particular instanton orbit with Pontryagin 
number $\k[A]=1$. In section~4 we construct this space explicitly in
Landau gauge. On the other hand, we see in section~5 that the Euler
character of the moduli space $\EE_\FF[A]$ vanishes in the
topologically trivial sector $\k[A]=0$ for gauges\equ{background}
defined by any nondegenerate background $B$. These results show
that global characteristics of the gauge theory can depend on the
gauge fixing condition employed. We conclude by discussing the
possible relevance of this dependence for the quantization of a gauge
theory.

\section{Gauge fixing and TQFT}
Gauge fixing an $SU(n)$ YM-theory with
the gauge condition\equ{cond} amounts to the statement that the
partition function 
\eq
\ZZ [A]=\int [dU][dc][d\cb][db]\ \exp S_{A }  
\eqn{gentqft}
with the action 
\be\label{gaugeaction}
S_A &=&- 2\int d\! x  s \tr \left(\cb(x)  \FF(A^U)(x)\right)\nn\\
  &=& - 2\int d\! x \tr \left(b(x)\FF(A^U)(x) -
\int d\! y \cb(x)\fud{\FF(A^U)(x)}{U(y)} c(y) \right)   
\ee
does not vanish and is independent of the orbit which the connection $A$
represents. In this case, the partition
function\equ{gentqft} 
could be inserted in the gauge-invariant measure of the YM-theory and
the  change of variables $A^U\rightarrow A$ would yield
the usual gauge-fixed action proposed by Faddeev and Popov. The so
far rather formal functional integral in\equ{gentqft} extends
over the gauge-group elements $U(x)\in SU(n)$, the Lie-algebra
valued Nakanishi-Lautrup
field $b(x)$ as well as the Lie-algebra valued anticommuting  Faddeev-Popov
ghosts and anti-ghosts $c(x)$, $\cb(x)$. The functional derivative
in\equ{gaugeaction} 
is computed with respect to right multiplication of the group.  

In ref.\cite{us96} it was observed that\equ{gentqft} with
action\equ{gaugeaction} can be regarded as the partition function of
a TQFT in the gauge group with the BRST algebra,
\pagebreak
\be 
\label{top}
sU(x) &=& U(x) c(x) \nn \\
sc(x) &=& -\half [c(x),\,c(x)] \nn \\
s\cb(x) &=& b(x)\nn \\
sb(x)&=&0
\ee
Note that the connection $A$ is a parameter
 of the TQFT  ($s A=0$). The nilpotent BRST algebra\equ{top} 
suggests\cite{bonora} that one may identify the ghost fields as
Maurer-Cartan forms on the infinite dimensional space of gauge
transformations. This geometrical interpretation of $c(x)$ as a
basis for the cotangent space at $U$ becomes particularly convincing
in the vicinity of $U=\one$.
The BRST operator can be identified
as the coboundary operator of the Lie algebra cohomology and
equations\equ{top} are regarded as the explicit representation of
its action at an arbitrary point $U(x)$ in the space of gauge
transformations. General arguments (for a review see e.g. \cite{Rak})
suggest, that the TQFT with the  $s$-exact action\equ{gaugeaction}
computes the Euler characteristic of the space of gauge transformations.   
One however first has to make sense of the Euler
character of an infinite dimensional space.

There are essentially two ways to compute the Euler characteristic of a
manifold. One is by use of the  
Gauss-Bonnet theorem, which gives the Euler characteristic as
an integral of the Euler class over the manifold. The other makes
use of the Poincar{\'e}-Hopf theorem, which amounts to counting with signs the
number of isolated zeros of some vector field $\FF=\vec{\nabla}V $ generated by
a potential $V$ on the manifold. The  signs are
determined by the  
sign of the Hessian at these isolated points. More generally, when the 
zero locus of the vector field $\FF $ is finite dimensional, the
regularized Euler characteristic of a manifold $\MM$ is 
\eq
\chi(\MM)=\sum_{X: \FF|_{X}=0}(\pm 1)\chi(X)
\eqn{rec}
where the signs depend on the orientation of the
fixed point manifolds $X$ embedded in $M$. The latter method can be
generalized to the case of 
infinite dimensional manifolds,  where  equation\equ{rec} is regarded
as the {\em definition}  of the regularized Euler characteristic of
the manifold\cite{AG}. 

The usual argument that the TQFT with BRST-algebra\equ{top} computes
the right hand side of\equ{rec} is based on the fact that the saddle
point approximation is apparently an exact evaluation of the functional
integral.  To see this one can for example modify the
action\equ{gaugeaction} by the $s-$exact term ${\a\ov 2} b^2$: 
\be\label{agauge}
S'_A &=&- 2\int d\! x  s \tr \left(\cb(x)\lf( \FF(A^U)- {\a\ov
2}b(x)\rt)\right)\nn\\ 
  &=& 2\int d\! x \tr \left ({\a\ov 2}b^2(x) - b \pa\ipr A^U +\int d\! y
\cb(x)\fud{\FF(A^U)(x)}{U(y)} c(y) \right)    
\ee
Due to the topological nature of the theory, the partition function
should be independent of $\a$. Gaussian integration over the multiplier
field $b$ in the functional integral with action $S'_A$ shows that
\eq
\ZZ [A]\propto \int [dU][dc][d\cb] \exp\lf(2\int d\! x\lf(-{1\ov
2\a}(\FF(A^U))^{2}+ \int d\! y \cb(x)\fud{\FF(A^U)(x)}{U(y)} c(y)
\right)\rt)    
\eqn{pfag}
In the $\a\rightarrow 0$ limit only fluctuations  around $U$'s
satisfying the fixed point equation $\FF[A^U]=0$ will
contribute to the functional integral. With the correct
(alpha-dependent) normalization the contribution of an isolated fixed
point to the partition function\equ{gentqft} 
 is  $\pm 1$ depending on the sign of the ``Hessian'' at that
point. To evaluate the contribution from a finite dimensional subspace
of fixed points, one introduces local coordinates and restricts the BRST-algebra\equ{top} to
that space. This procedure generally induces curvature
terms\cite{blau}.  One may then  use the Gauss-Bonnet theorem to find that the
contribution to the partition function of the TQFT  is the Euler
characteristic of the submanifold\cite{thom}. In the $\alpha\rightarrow 0$
limit, the partition 
function of the TQFT is thus seen to reproduce the right hand side
of\equ{rec}.

It could appear that the  space of gauge-transformations and therefore
its Euler characteristic can be defined independently of the
connection $A$ and that the TQFT\equ{gentqft}
does not depend on $A$ at all. Let us stress however, that the Euler
characteristic computed via the Poincar{\'e}-Hopf theorem is actually
that  of the {\it domain} $D:=\{x\in\MM, |V(x)|<\infty\}$, rather than
the Euler characteristic of the whole
manifold $\MM$. Thus the Euler characteristic
defined by\equ{rec} coincides with that of the original manifold
$\MM$ only if the potential is finite everywhere. 

If the TQFT on the gauge group  is to be employed as a gauge fixing
device, the gauge condition\equ{cond} will only  depend on the orbit
$\{A^U\}$ rather than $U$ itself. Any background gauge condition of the
form\equ{background} is the gradient of an associated potential,
introduced in\cite{Zw}\footnote{In
contrast to\cite{Zw} we here however do not select a 
representative connection by the absolute minimum of the
potential\equ{pot} on the gauge orbit, but rather sum over all its
relative minima in the sense of the Poincar{\'e}-Hopf theorem.}  
\eq
V[U]=\norm{A^U-B}^2=\half\int d\! x\
Tr(A^{U}-B)\ipr(A^{U}-B)=\half\int d\! x\
Tr(A-B^{U^\dagger})\ipr(A-B^{U^\dagger}) 
\eqn{pot}
Indeed, the fixed points of this potential $\d V[U]=0$
\[
{\d V\ov\d U(x)}= -\FF(A^U)(x)
\]
are just the solutions to the gauge condition\equ{background}. 
The domain of the potential\equ{pot} in the sense of the Poincar{\'e}-Hopf
theorem consists only of 
those gauge transformations $\{U\}$ for which the connection $A^U-B$
is square integrable. This domain does not depend on continuous
deformations of the connections $A$ and $B$ but may (and we will
indirectly see that it does) depend on their topological characteristics. 

Completing the space of $C^\infty$ connections in the $L^2$ norm
$\norm{\cdot}$ was shown\cite{DAZ} to naturally
extend to considering the space of $C^\infty$ gauge
transformations completed in the Sobolev norm $\norm{U}_1=\norm{U} +
\norm{\pa U}$. For the TQFT it is important that this is a topological
space of gauge transformations which furthermore fully describes the
gauge orbit $\ol{\{A^U\}}=\{A^{\ol{\{U\}}}\}$\cite{DAZ}. We know of no
other space of gauge transformations where this important property
holds and will therefore  work in this space. Note that in order
to span the space of gauge transformations in the
neighbourhood of a particular $U(x)$ one has to consider all
fluctuations $\d U$  which are normalizable in the $L^2$-norm
$\norm{\d U}$. To preserve invariance under infinitesimal isometry
transformations  $x\rightarrow 
x+\e(x)$ of  the base manifold,  $U(x)$ as well as
$U(x+\e(x))$ have to belong to the space of allowed gauge
transformations. This implies that one has to account for fluctuations
$\d U=U(x+\e(x))-U(x)$, which in general are only 
normalizable in the $L^2$-norm $\norm{\cdot}=\norm{\cdot}_0$, if $U$ is normalizable in
$\norm{\cdot}_1$. These considerations determine the
functional space we should consider in the TQFT with
BRST-algebra\equ{top}. 
\begin{itemize}
\item $U(x)\in C^\infty$ completed in the norm $\norm{\cdot}_1$
\item $b(x), c(x)$ and $\cb(x)$ in $C^\infty$ completed in the norm
$\norm{\cdot}=\norm{\cdot}_0$.
\end{itemize} 

Our previous argument indicates that the {\it regularized} Euler
characteristic so obtained does not necessarily coincide with the Euler
characteristic of the full space of gauge transformations and may depend on
topological properties of the gauge 
condition\equ{cond}. The purpose of this paper is to investigate this
possibility. 

For certain gauge conditions\equ{cond} it is easy to see that 
the regularized Euler characteristic computed in this way vanishes. This is
for instance the case, whenever the Euler characteristic of
each subspace of fixed points vanishes individually due to a group
isometry. One encounters
this situation in any background gauge\equ{background} defined
by a degenerate orbit $B$ and in particular in covariant gauges, which
correspond to choosing the degenerate background orbit $B=0$. The
associated potential $V$ is invariant with respect to
right multiplication by a certain group in this case and the fixed point spaces
therefore posess  an isometry generated by the group action. Since
this isometry has no fixed points, the Euler characteristic of each subspace
vanishes.   

The problem can be circumvented by an equivariant
BRST construction\cite{us96} which divides out this  group
manifold. In covariant gauges one considers $X/SU(n)$ rather than the space of
fixed points $X$ itself and the TQFT based on the equivariant
cohomology computes the Euler characteristic $\chi(X/SU(n))$.
For details on the equivariant BRST construction in covariant gauges
and the associated TQFT we refer to\cite{us96}. 

\section{Instantons on $S_4$ in covariant gauges}
The main reason for employing covariant gauges in Minkowski or
Euclidean space-time  is that they allow a manifestly relativistically
invariant formulation of the QFT. The natural generalization of this
distinction to other  spacetime manifolds like $S_4$, is that a
covariant gauge condition preserves {\it all} isometries of
the base manifold. Covariant gauges preserve the homogeneity and
isotropy of space-time and do not select a preferred
point and/or direction in the gauge-fixed theory. This could be
particularly important for defining the thermodynamic limit
unambiguously, especially since color forces are expected to be
strong and of infinite range.

By formulating a massless theory such as unbroken YM on
compact Euclidean space-time, one can avoid infrared divergences
while preserving the gauge- (or rather BRST-) symmetry. One hopes
that the thermodynamic limit of physical correlation functions can be
obtained by rescaling the compact base-manifold and that this
limit is independent of the manifold  used to formulate the
theory. The thermodynamic limit, if it exists, can at most depend on
topological characteristics of the base manifold. 

In the case of Yang-Mills theory on  compact
space-time with the topology of an $S_4$ it is well known that the
space of gauge orbits is disconnected. The possible importance  of sectors with
Pontryagin number $\k[A]\neq 0$ has  been recognized long ago\cite{Pagels} and
`t~Hooft's semiclassical calculation\cite{tH} indicates that this
could resolve the $U_A(1)$ problem\cite{We}. We wish to stress that
the $U_A(1)$-problem is intimately related to the existence of
covariant gauges, since one makes use of the Goldstone theorem to
formulate it\cite{We}. Due to an anomalous 
contribution,  the conserved $U_A(1)$ current is gauge dependent. In
covariant gauges the Ward identities nevertheless imply the existence
of a Goldstone pole in correlation functions of this current with
quark multilinears, if the $U_A(1)$-symmetry is spontaneously
broken\cite{colm}. It is therefore
important to verify that gauge orbits with Pontryagin number $\k[A]\neq
0$ contribute to physical correlation functions in {\it
covariant} gauges.

We adopt conformal coordinates to parametrize the $S_4$ and use its
diameter $2R=1$ as the unit of length. In these coordinates the
metric is diagonal ($x^2=\sum_\m x_\m x_\m$)
\eq
g_\mn (x) =g(x) \d_\mn = (1+x^2)^{-2} \d_\mn
\eqn{metric} 
and the invariant volume element of the $S_4$ is simply $dx=d^4x g^2(x)$.

The SO(5) isometry group of an $S_4$ is generated by 
the  coordinate transformations
\eq
x^\prime_\m =x_\m -\epsilon_\m (x;a,\omega)= x_\m -\omega_\mn
x_\n + a_\m (x^2-1) -2 x_\m x\ipr a 
\eqn{isos}
depending on ten infinitesimal parameters $a_\m$ and
$\omega_\mn=-\omega_{\n\m}$. 

A scalar density $s(x)$ transforms as
\eq
\delta s(x)= s_{,\m}(x) \epsilon_\m(x; a,\omega)
\eqn{scalar}
 under the isometry group.
Together with the variation of vector
fields such as a connection $A_\m(x)$
\eq
\delta A_\m(x) = A_{\m,\n}\epsilon_\n(x;a,\omega) + A_\n(x)
\epsilon_{\n,\m}(x;a,\omega)  
\eqn{vector}
\equ{scalar} determines the transformation properties of all higher
rank tensors. The
isometry generators\equ{isos} of course do not change the conformal
metric\equ{metric} 
\eq
\delta g_\mn(x) =0=g(x) (\epsilon_{\m,\n}(x; a,\o)
+\epsilon_{\n,\m}(x; a,\o) ) + \delta_{\mn}
g_{,\s}(x)\epsilon_\s(x; a,\o)
\eqn{isometric}
The change of a pseudoscalar density $p(x)$ such as the YM-Lagrangian
\eq
\LL(x)=\frac{1}{2 g^2}\sqrt{det(g_\mn)}\, \tr F_\mn F_{\r\s}
g^{-1}_{\m\r} g^{-1}_{\n\s},  
\eqn{lagrangian}
or the  Pontryagin density
\eq
\PP(x)=\frac{1}{32\pi^2}\e_{\mn\r\s} \tr F_\mn F_{\r\s}
\eqn{pontdensity}
under conformal transformations
\eq
\delta p(x)= p_{,\m}(x) \epsilon_\m(x; a,\omega) + p(x)
\epsilon_{\m,\m}(x; a,\omega) 
\eqn{pseudo}
will also be useful in the following.

The prototype of a covariant gauge condition
is the  Landau gauge, which on an $S_4$ in conformal coordinates takes
the form
\eq
\FF(A)=\pa\ipr A =(\det g)^{-\half} \pa_\m (\det g)^\half g^{-1}_\mn A_\n(x) =
g^{-2}(x) \pa_\m g(x) A_\m(x)=0
\eqn{landau}
The following considerations are however not particular to  Landau gauge
and only make use of the fact that the
gauge condition $\FF(A)$ is covariant. To investigate topological
characteristics of the space of
solutions\equ{ef} to a covariant gauge condition in the sector with
Pontryagin number $\kappa=1$, we will choose a
particular orbit in that sector. 

As discussed in the previous section the space of solutions\equ{ef} to
a covariant gauge condition like\equ{landau} for general orbits only
posesses an isometry with respect to right multiplication by constant
gauge group elements. The space of solutions\equ{ef} of certain orbits
can however have a larger isometry. Covariant gauges are invariant
under the $SO(5)$ isometry group of the $S_4$ by definition, and we
will find a particular connections $A^{(s)}$ in the $\k=1$ sector
for which isometry transformations\equ{vector} of the
base space are equivalent to infinitesimal gauge transformations  
\eq
\d A^{(s)}_\m(x) = A^{(s)}_{\m,\n}\epsilon_\n(x;a,\omega) + A^{(s)}_\n(x)
\epsilon_{\n,\m}(x;a,\omega) = D^{A^{(s)}}\theta(x;a,\omega)
\eqn{equivalence}
The  equivalence\equ{equivalence} between isometry- and
gauge- transformations is only possible for an orbit whose classical
YM Lagrangian\equ{lagrangian}
and Pontryagin density\equ{pontdensity}
are both invariant under the  $SO(5)$ isometry group. The
$SO(4)$-invariance of $\LL(x)$ and $\PP(x)$ 
implies that these pseudoscalar densities  are only functions of $t=x^2$. Using
\equ{pseudo} and \equ{isos} the required invariance with respect to
the $SO(5)/SO(4)$ coset generators gives the differential equation
\eq
[(1+t)\frac{d}{dt} + 4]\left\{ \begin{array}{c}\PP(t)\\ \LL(t)\end{array}\right. = 0\ ,
\eqn{instantoneq}
determining $\PP(t)$ and $\LL(t)$ up to a
normalization. The normalization of $\PP(t)$ is fixed by the
Pontryagin number $\kappa=1$ and  for $\LL(x)$ can be absorbed 
in the definition of the coupling constant. The solution of
\equ{instantoneq} is  thus seen to be just
the familiar pseudoscalar density of  an BPST-instanton\cite{BPST} with scale
$\rho=2R=1$ located at the ``origin'' $x=0$
\eq
\frac{g^2}{8\pi^2}\LL(x)=\PP(x)=\frac{6}{\pi^2} (1+x^2)^{-4} = {\frac 1
{32\pi^2}}\epsilon_{\mn\r\s}\tr F^{(s)}_\mn F^{(s)}_{\r\s}  
\eqn{FFdualdensity}

A selfdual configuration with
Pontryagin number $\k=1$ will be called a {\it standard} instanton in
what follows if its Pontryagin density is invariant under the SO(5)
isometry group of the $S_4$. In conformal coordinates, its Pontryagin
density is given by\equ{FFdualdensity}. 

Jackiw and Rebbi\cite{Ja76} studied the BPST-instanton\cite{BPST}
solutions to the classical 
equations of motion of an $SU(2)$ Yang-Mills theory on an Euclidean
$S_4$. The space of these solutions
forms an $SO(5,1)/SO(5)$ modulo gauge transformations, which tallies
with the fact that the moduli-space in the $\k=\pm1$ sector of an
SU(2)-theory  is 5~dimensional\cite{AH}. They 
found  that a particular connection of the standard instanton is 
invariant under the $SO(5)$ isometry group of the $S_4$ modulo gauge
transformations.  Actually this  does not depend on the point on the
gauge orbit of the standard instanton, but is true on the whole orbit. 
This slight generalization of their result enables us to
also study the infinitely many Gribov copies of the connection
considered in~\cite{Ja76} without the need to explicitely construct them. 

The generalization is possible because the moduli-space of an $SU(2)$
instanton is only 5~dimensional\cite{AH} and very well known. Any
variation of  $A^{(s)}$ which  does not 
change the YM action is either a gauge transformation or would have
to dilate or translate the standard instanton and therefore show up as
a change of its Pontryagin  density. There are no {\it hidden}
moduli parameters in the $\k=1$ sector on which the Pontryagin 
density of an instanton does not depend. Since the Pontryagin density
of a standard instanton is invariant  under isometry transformations, we are
assured that this transformation does not move in the
moduli-space of the instanton and must be a gauge transformation
only. The equivalence\equ{equivalence} therefore holds at 
every point on the gauge orbit of a standard instanton.  

We still have to determine the space these gauge modes of the
standard instanton actually span. Only the trivial configuration
$A=0$, is invariant under the full $SO(5)$ isometry 
group. Vector fields which are
invariant under an $SO(4)$ subgroup of $SO(5)$ are pure gauge -- the
corresponding anti-symmetric field strength tensor, an $(1,0)\oplus(0,1)$
representation of $SO(4)$,  has to vanish since only the null vector
is invariant under $SO(3)$-rotations\footnote{A more 
pedestrian proof of this statement is obtained by using conformal
coordinates adapted to the $SO(4)$ in question. The
invariant vector field in these coordinates is of the form $A_\m(x)
= x_\m \Phi(x^2)$ and easily seen to be pure gauge.}. The best one
can achieve in the $\k\neq 0$ sectors is invariance of the connection
under an $SO(3)$ subgroup of $SO(5)$.  The corresponding field
strength tensor is then (anti-)selfdual. We therefore conclude that:

{\it A BPST-instanton in the $\k=1$ topological sector is changed by
(or breaks) the 7 generators of the $SO(5)/SO(3)$ coset space of the
isometry group of an $S_4$. The broken generators form 
an $(\half\,,\,\half) \oplus (1\,,\,0)$ representation of $SO(4)$ } 

We already know that isometry transformations of the {\it
standard} instanton are equivalent to gauge transformations.  The
isometries of an $SO(5)/SO(3)$ coset space therefore correspond
to {\it non-trivial} gauge transformations of the standard instanton. The
three broken  generators of the $(1\,,\,0)$ representation obviously
generate constant gauge transformations. The other four broken generators
in the $(\half\,,\, \half)$ representation must generate gauge
modes of the standard instanton which are {\it linearly independent}
of these  -- simply because they transform according to different
representations of the $SO(4)$ subgroup. 

To summarize: {\it isometry transformations  of an $S_4$ are mapped 
to an $SO(5)/SO(3)\simeq S_7$ subspace of the automorphisms of the gauge orbit
of a standard instanton. Modulo global $SU(2)$ transformations, the
space of solutions to a covariant gauge condition on the orbit of a
standard instanton has the topological structure} 
\eq 
\EE_{cov}[A^{(s)}]\simeq \frac{SO(5)}{SO(3)\times
SU(2)}\times\BB\simeq S_4\times \BB
\eqn{solspace}
The space $\BB$ was not determined and could depend on the
covariant gauge condition employed. The result\equ{solspace} shows
that the gauge orbit of a standard instanton is {\it on} the Gribov
horizon in any covariant gauge.   

Our arguments did not make use of  any 
{\it explicit} form for the connection of a  standard instanton. They only
rely on the special property\equ{equivalence}  of the orbit of a
standard instanton and apply to any covariant gauge.

\section{The Euler characteristic of the space of solutions to
covariant gauge conditions in the $\k=\pm 1$ sectors}
In Landau gauge two explicit standard instanton connections have been
studied extensively. In conformal coordinates on an $S_4$ they are 
\eq
A_\m^{(1)}(x) = \frac{x^2\ u^\dagger\pa_\m u}{1+x^2} =\frac{2\eta_\mn
x_\n}{1+x^2}     
\quad{\rm and}\quad A_\m^{(2)}(x)=A_\m^{(1)u^\dagger}(x) =
\frac{u\pa_\m u^\dagger}{1+x^2}=\frac{2 \bar\eta_\mn x_\n}{x^2(1+x^2)}
\eqn{standard12}
$(\bar\eta_\mn),\eta_\mn $ are the Lie-algebra valued
(anti-)selfdual tensors  introduced by `t Hooft\cite{tH}. As
indicated in\equ{standard12} $u(x)=(x^4\one +i\vec\s\vec
x)/\sqrt{x^2}$  is the gauge transformation relating these two
connections. Both connections\equ{standard12} have finite $L^2$-norm
$\norm{\cdot}$ and the gauge transformation $u(x)$ relating them is
normalizable in $\norm{\cdot}_1$. In the functional space we are
considering, the connections\equ{standard12} therefore belong to the
{\it same} gauge orbit and are  {\it both} instantons with Pontryagin
number $\k=+1$. 

In fact they are just two points of an $S_4$ of gauge equivalent
instantons parameterized by $b\in\RR^4$ 
\eq
A_\m^{(b)}(x)= A_\m^{(1)\, u^\dagger(x+b)}(x)=u(x+b)
A_\m^{(1)}(x) u^\dagger(x+b) + u(x+b)\pa_\m u^\dagger(x+b)
\eqn{s4}
Using
\eq
u^\dagger (x)\pa_\m u(x)=\frac{ 2\eta_\mn x_\n}{x^2}
\eqn{udu}
and the $su(2)$-algebra of the selfdual tensors $\eta_\mn$
it is straightforward to verify that any connection\equ{s4} satisfies
the Landau gauge condition\equ{landau}. Modulo constant gauge
transformations $ A^{(b\rightarrow\infty)}\simeq A^{(1)}$
and $A^{(b=0)}=A^{(2)}$. Points at infinity in the parameter space correspond to connections $A^{(b)}$ which are equivalent modulo constant gauge
transformations and are identified in equivariant TQFT.
 Thus the parameters $b$ can be
considered projective coordinates on  $S_4$ of gauge equivalent
standard instanton connections satisfying the Landau gauge
condition. The connections\equ{standard12} are those at the ``north'' and
``south'' poles of this $S_4$. One may also verify explicitely that an
infinitesimal variation of the parameters $b$ in $A^{(b)}$ corresponds
to an isometry transformation\equ{isos}. 

The construction in Landau gauge  shows that there is no
further identification of connections on the $S_4$ since {\it all}
constant gauge transforms of $A^{(1)}$ are identified with a single
point on the $S_4$. The space of solutions to the
Landau gauge condition on the orbit of a standard instanton therefore
indeed has the structure\equ{solspace} and we conclude from
$\chi(S_4)=2$ that 
\eq
\chi(\EE_{\pa\ipr A}[A^{(s)}])= even
\eqn{eulerlandau}
The Euler characteristic of the space should not depend on
continuous deformations of the orbit or the gauge condition. Hence for a 
generic {\it covariant} gauge
condition which can be continuously deformed to Landau gauge , the Euler 
characteristic of the space of solutions on
orbits in the $\k=\pm 1$ sectors  is {\it even}
\eq
\chi_{cov}(\k[A]=\pm 1)= even
\eqn{eulercov1}
To determine the Euler characteristic more accurately than in\equ{eulercov1}
would require a better understanding of the so far undetermined
space $\BB$ in\equ{solspace}.
Together with the previous result of ref.\cite{us96} for the $\k=0$
sector  on $S_4$
\eq
\chi_{cov}(\k[A]=0)= 1
\eqn{eulercov0}
\equ{eulercov1} however already shows that the partition function of
the TQFT {\it depends} on the topological sector of the orbit in
covariant gauges. 

\section{General background gauges and possible 
physical implications of global gauge dependence}

In the previous section we observed that the partition function of the
TQFT associated with a covariant gauge fixing condition depends on the
topological sector of the orbit. Although
we cannot construct the space of solutions for an arbitrary background
gauge\equ{background}, we will see that its Euler characteristic
vanishes for orbits in the trivial topological sector $\k=0$ when the
background connection $B$ is not degenerate. Thus the partition
function of the TQFT either vanishes identically in this case, or
 depends on the topological sector also in these gauges.

Consider for instance a gauge condition\equ{background} where the
background connection $B$  is 
the instanton connection $A^{(1)}$. Since the orbit of
$B$ is non-degenerate\footnote{$D^B\o=0$ only has the trivial solution
$\o=0$ when $B=A^{(1)}$ of\equ{standard12}}, the associated
potential\equ{pot} in this case has a unique absolute minimum at
$U=\one$ for $A=B$. There is no degenerate subspace of solutions to
divide out in this gauge, and the equivariant construction of
ref.\cite{us96} cannot be employed.  'T Hooft used this gauge for his
semiclassical calculation\cite{tH}.  The Faddeev-Popov
operator in this case is $D^{B}\ipr D^{A^U}$. At the absolute minimum of the
potential at $A^U=B=A^{(1)}$ this operator is positive definite. Although
we don't have much  control over the Gribov copies in this gauge, the
$\k=1$ sector may very well contribute to the partition function of
the TQFT.

On the other hand this background gauge condition is degenerate for
the flat orbit $A=U^\dagger dU$. The Faddev-Popov operator $D^B\ipr
D^{U^\dagger d U} $ on this orbit apparently has zero-modes
corresponding to left multiplication of $U$ by global gauge 
transformations and the space of solutions to this gauge condition
on the flat orbit has the topological structure 
\eq
{\cal E}_{D^B\ipr A}[A=0]=\{U:D^{B}(U^{\dagger}\pa U)=0\}\simeq
SU(n)\times\bar \BB.
\eqn{insgauge}
Since the Euler character of a group manifold vanishes, the Euler
characteristic of\equ{insgauge} vanishes irrespective of the space
$\bar \BB$. We thus see that orbits in the $\k=0$ sector cannot
contribute to the  partition function of the TQFT in this gauge. The
above argument is easily extended to any non-degenerate background
connection $B$ -- the space of solutions to the background gauge
condition on the orbit of flat connections has vanishing Euler
characteristic in this case and it is impossible to remove this
degeneracy with an equivariant construction analogous to that
of\cite{us96}.

We have shown for an $S_4$ that an orbit with Pontryagin number
$\k=\pm 1$ contributes to physical correlation functions in
covariant gauges with a different multiplicity factor due to Gribov copies,
 than an orbit in the trivial
topological sector and that nondegenerate background gauges generally
annihilate the $\k=0$  sector of the theory altogether.   
This confirms the claim of ref.\cite{us96} that 
global properties of the {\it quantized} gauge theory may depend on
the gauge fixing.  Global properties of a YM-theory  therefore will in general
depend on the gauge\footnote{Since perturbation theory only accounts
for fluctuations around a {\it single} solution to the gauge
condition, it is not sensitive to the global issues discussed here.}.

Although this ambiguity
was  found in the continuum formulation of a gauge fixed YM-theory, it
could also arise in the thermodynamic limit of a
lattice gauge formulation. From this point of view it is perhaps less
surprising that topologically nontrivial configurations contribute
mainly in lattice gauge theories with non-covariant boundary conditions.

Our observation can have  implications
for the $U_A(1)$- and strong CP-violation  problem, if it turns out
that topologically non-trivial sectors do not contribute in
covariant gauges. Our topological  arguments are not sufficiently
refined to actually determine the regularized Euler character in the
$\k\neq 0$ sectors of the theory in covariant gauges. They only
indicate that it is {\it even} and therefore differs from that of the
$\k=0$ sector. It is obviously desireable to 
improve on this result. Our investigation nevertheless shows that the well
known global dependence of a  YM-theory on the 
topology of  compact space-time may also depend on the gauge
condition. 

Physical correlation functions are gauge
invariant and any gauge dependent answer is usually attributed to a badly
performed gauge fixing. This however presumes that the quantization of
a classical gauge theory is unique. There may  be
many  different quantized gauge theories which 
correspond to a single classical theory with a certain {\it
infinitesimal} gauge invariance.  There is
no {\em a priori} reason that so quantized ``gauge theories'' are
identical, since the extension of this procedure to the whole orbit space
will not be unique in general.
Of course, only {\it one} of these extensions can (at best) describe physical
reality and physical criteria have to be used to select this
model. The phenomenon is very similar to
the spontaneous breakdown of a symmetry where the realistic model is
selected by choosing one of the degenerate vacua.

Our investigation only addressed theories with a 
BRST-symmetry and not the orbit space of gauge theory {\em per se}. We
adopted  BRST symmetry as the 
guiding principle for constructing a gauge theory. The relation of the usual
gauge fixing procedure to a certain TQFT in the gauge group observed
in\cite{us96} allowed us to determine global characteristics of
such a theory. We believe that our definition of partition function based on a complete BRST approach will allow  to investigate the
Gribov problem in more details. We speculate that physical properties such as 
the apparent absence of strong CP-violation may select a preferred class
of ``gauges''.  

\vspace{.5cm}
\begin{center}
\bf Acknowledgements
\end{center}\nobreak
We would like to thank M.~Porrati and D.~Zwanziger for illuminating
and very helpful discussions.

}\end{document}